\documentclass[aps,prl,twocolumn]{revtex4}
\usepackage{amsfonts}
\usepackage{amsmath}
\usepackage{graphicx}
\usepackage{dsfont}
\usepackage{diagbox}
\usepackage{array}
\usepackage{amssymb}
\usepackage{bbm}
\usepackage{float}
\usepackage{url}
\usepackage{hyperref}
\usepackage{booktabs}
\usepackage{color}

\begin{document}

\title{Phase coherence of pairs of Cooper pairs as quasi-long-range order of
half-vortex pairs in a two-dimensional bilayer system}
\author{Feng-Feng Song$^{1}$ and Guang-Ming Zhang$^{1,2}$}
\email{gmzhang@tsinghua.edu.cn}
\affiliation{$^{1}$State Key Laboratory of Low-Dimensional Quantum Physics and Department
of Physics, Tsinghua University, Beijing 100084, China. \\
$^{2}$Frontier Science Center for Quantum Information, Beijing 100084, China.}
\date{\today}

\begin{abstract}
It is known that the loss of phase coherence of Cooper pairs in
two-dimensional (2D) superconductivity corresponds to the unbinding of
vortex-antivortex pairs with the quasi-long-range order (quasi-LRO) in the
order-parameter phase field, described by the
Berezinskii-Kosterlizt-Thouless (BKT) transition of a 2D XY model. Here we
show that the second-order Josephson coupling can induce an exotic
superconducting phase in a bilayer system. By using tensor-network methods,
the partition function of the 2D classical model is expressed as a product
of 1D quantum transfer operator, whose eigen-equation can be solved by an
algorithm of matrix product states rigorously. From the singularity shown by
the entanglement entropy of the 1D quantum analogue, various phase
transitions can be accurately determined. Below the BKT phase transition, an
inter-layer Ising long-range order is established at $T_{Ising}$, and the
phase coherence of both intra-layers and inter-layers is locked together.
For two identical layers, the Ising transition coincides with the BKT
transition at a multi-critical point. For two inequivalent layers, however,
there emerges an intermediate quasi-LRO phase ($T_{Ising}<T<T_{BKT}$), where
the vortex-antivortex bindings occur in the layer with the larger
intra-layer coupling, but only half-vortex pairs with topological strings
exist in the other layer, corresponding to the phase coherence of pairs of
Cooper pairs. So our study provides a promising way to realize the charge-4e
superconductivity in a bilayer system.
\end{abstract}

\maketitle

\textit{Introduction.} -Superconductivity arises from electron pairing and
its phase coherence. In conventional Bardeen-Cooper-Schrieffer
superconductors, the electron pairing and condensation of Cooper pairs
always happen simultaneously, and the superconducting transition is
determined by the pairing temperature. In two dimensions (2D), however, the
true transition can be substantially below the pairing temperature and is
controlled primarily by thermal fluctuations in the phase field of the order
parameter\cite%
{Beasley_1979,Emery-Kivelson,Carlson-1999,Li-2007,Ding-2008,Wu-2020,Shen-2021}%
. In the Ginzburg-Landau theory, when the magnitude fluctuation of the order
parameter is frozen, the phase field fluctuation can be characterized by the
2D XY spin model, and the loss of phase coherence among the Cooper pairs
corresponds to the unbinding of vortex-antivortex pairs with the
quasi-long-range order (quasi-LRO), characterized by the
Berezinskii-Kosterlizt-Thouless (BKT) phase transition\cite%
{Berezinsky_1970,Kosterlitz_1973,Kosterlitz_1974}.

In recent years there has been the increasing interest in a bilayer
structure of coupled 2D superconducting systems\cite%
{Bojesen-2013,Bojesen_2014,Kobayashi-2019,Bighin-2019, Zeng-Wu-2021}. When a
direct Josephson coupling is present, the relative phase of the order
parameters is pinned to a fixed value, so both phase locking and phase
coherence of the Cooper pairs are characterized by a single BKT transition%
\cite{Parga-1980}. However, when the direct Josephson coupling is suppressed%
\cite{Li-2007,Ding-2008}, the second-order Josephson coupling is dominant,
and an Ising-like transition for the phase locking occurs at $T_{Ising}$,
which is usually lower than the BKT transition temperature $T_{BKT}$. For
the inequivalent coupled layers, it was argued the existence of an
intermediate phase ($T_{Ising}<T<T_{BKT}$) with partial order: one layer is
in disordered phase and the other layer have vortex-antivortex pairs with
quasi-LRO\cite{Grannato-1986}. Due to the lack of sharp thermodynamic
signatures for the BKT transition, it cannot unambiguously determine whether
the transition for the identical coupled layers is a single or double
transitions with an intervening unlocked phase\cite{Grannato-1986}.
Actually, the nature of the intermediate phase with partial order has not
been fully explored, so it is a great challenge to determine the global
phase diagram and calculate the properties of the intermediate phase
accurately.

Recently, tensor network methods have become a powerful tool to characterize
correlated quantum many-body phases and their phase transitions in the
thermodynamic limit\cite{Verstraete-Adv-Phys,Orus-Ann-Phys}. Since the
partition function of a 2D statistical model can be represented as a tensor
product of 1D quantum transfer operator\cite{Haegeman-Verstraete2017}, the
correspondence eigen-equation can be efficiently solved by the algorithm of
variational uniform matrix product states\cite%
{VUMPS,Fishman_2018,Laurens_Haegeman_2019,Laurens_Bram_2019}. In this
Letter, we apply this method to the bilayer system. According to the
singularity displayed by the entanglement entropy of the 1D quantum
analogue, various phase transitions can be precisely determined\cite%
{Li_2020,Song-Zhang-2021}, and various correlation functions of local
observables are calculated rigorously.

\begin{figure}[tbp]
\centering
\includegraphics[width=0.45\textwidth]{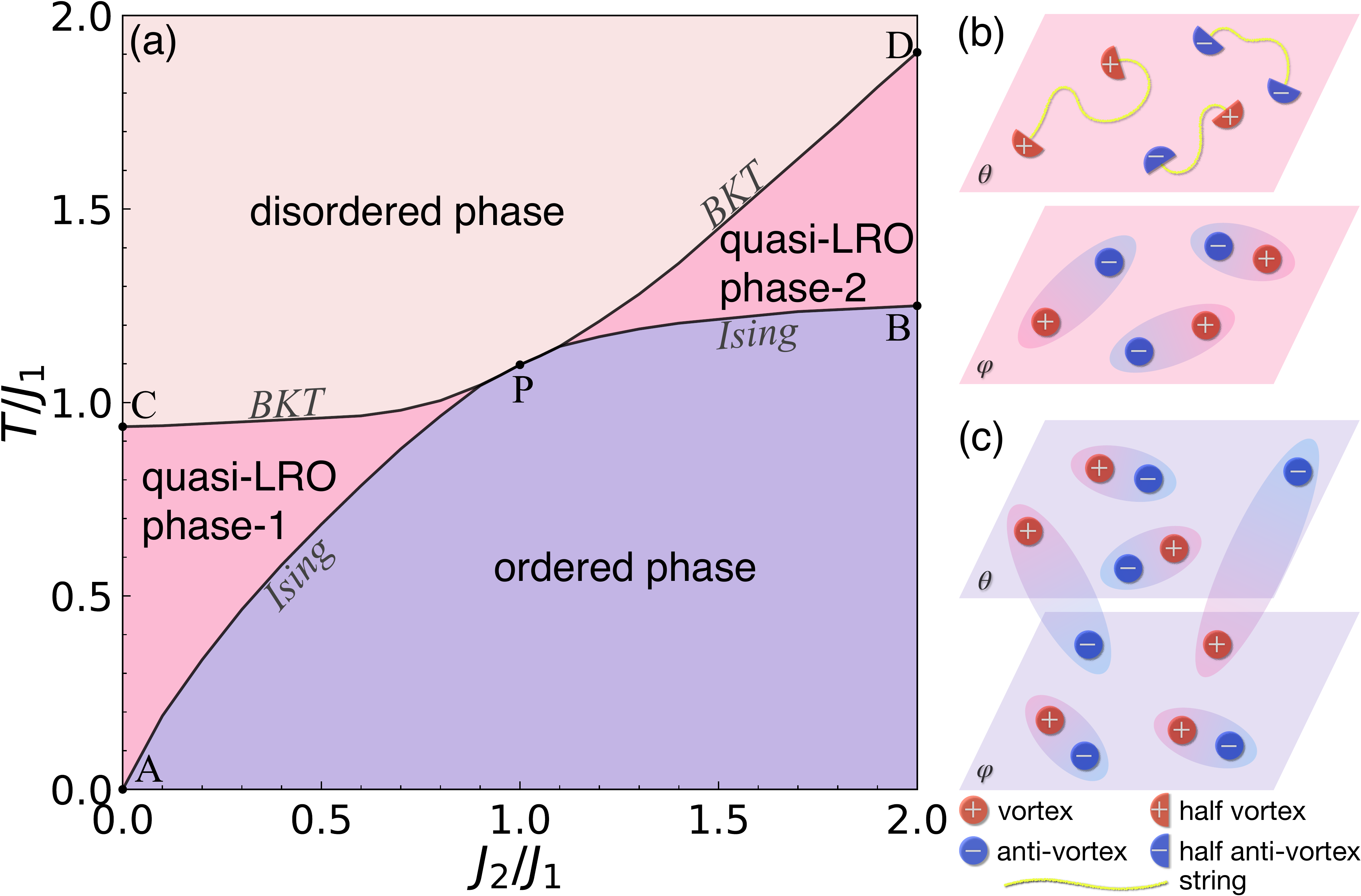}
\caption{ (a) The finite-temperature phase diagram of the bilayer system.
Here we choose $K=0.5J_1$. In the low temperature phase, there emerges an
inter-layer Ising-like long-range order. The BKT and Ising transitions merge
together at the point $P$. (b) The schematic picture of the quasi-LRO
phase-2, while the quasi-LRO phase-1 is obtained by switching the upper and
lower layers. (c) The schematic picture of the low-temperature ordered
phase. }
\label{fig:phase}
\end{figure}

The derived global phase diagram is displayed in Fig.\ref{fig:phase}(a). As
the temperature decreases, the BKT transition first occurs before a local
inter-layer Ising long-range order is established. The Ising transition
accompanies with the vortex-antivortex bindings in both intra-layers and
inter-layers, as shown in Fig.\ref{fig:phase}(c). For two identical layers,
the Ising transition coincides with the BKT transition at the multi-critical
point $P$. However, for two inequivalent layers, we find that the
intermediate phase has a quasi-LRO: vortex-antivortex bindings occur only in
the layer with the larger intra-layer coupling while half-vortex pairs
emerge in the other layer, schematically shown in Fig.\ref{fig:phase}(b).
Since the half-vortices are point singularities around which spin directions
rotate through an angle $\pi$ on circumnavigation, each pair of
half-vortices is connected by a topological string\cite%
{Lee_1985,Carpenter_1989,Shi_2011,Serna_2017}. More importantly, as the
quasi-LRO of the phase fields can be viewed as the condensation of the
Cooper pairs of 2D superconductivity, the half-vortex pairs with a quasi-LRO
imply the condensate of pairs of the Cooper pairs in the absence of phase
coherence among the Cooper pairs, corresponding to the charge-4e
superconductivity\cite{Doucot_2002,Babaev_2004,Berg_2009,Yao_2017,Fu_2021}.

\textit{Model and tensor-network methods}. -The Hamiltonian of a two-coupled
XY spin model on a square lattice is defined by
\begin{eqnarray}
H &=&-J_{1}\sum_{\langle i,j\rangle }\cos (\theta _{i}-\theta
_{j})-J_{2}\sum_{\langle i,j\rangle }\cos (\varphi _{i}-\varphi _{j})  \notag
\\
&&+K\sum_{i}\cos (2\theta _{i}-2\varphi _{i}),
\end{eqnarray}%
where $\theta _{i}$ and $\varphi _{i}\in \lbrack 0,2\pi ]$ are two $U(1)$
phase fields describing the pairing order-parameters on the upper and lower
layers, respectively, $J_{1}$ and $J_{2}$ are their respective
nearest-neighbour intra-layer couplings, and $K$ denotes the second-order
Josephson inter-layer coupling. Due to the nature of the low-temperature
phase, the inter-layer coupling is always relevant for \textit{any} finite
value of $K$, and the phase fields $\theta$ and $\varphi$ are no longer two
independent $U(1)$ variables. At low temperatures, the relative phase $%
\sigma_{i}\equiv\theta_{i}-\varphi_{i}$ is reduced to a $\mathbb{Z}_2$
variable, which can be explicitly displayed in the limit of $%
K\rightarrow\infty $, $\varphi _{i}=\theta _{i}+\pi s_{i}/2$ with $s_{i}=\pm
1$. The reduced Ising-XY coupled model was intensively studied by various
numerical methods\cite%
{Choi-1985,Granato_1991,Lee_1991,Li_1994,Nightingale_1995}.

\begin{figure}[tbp]
\centering
\includegraphics[width=0.45\textwidth]{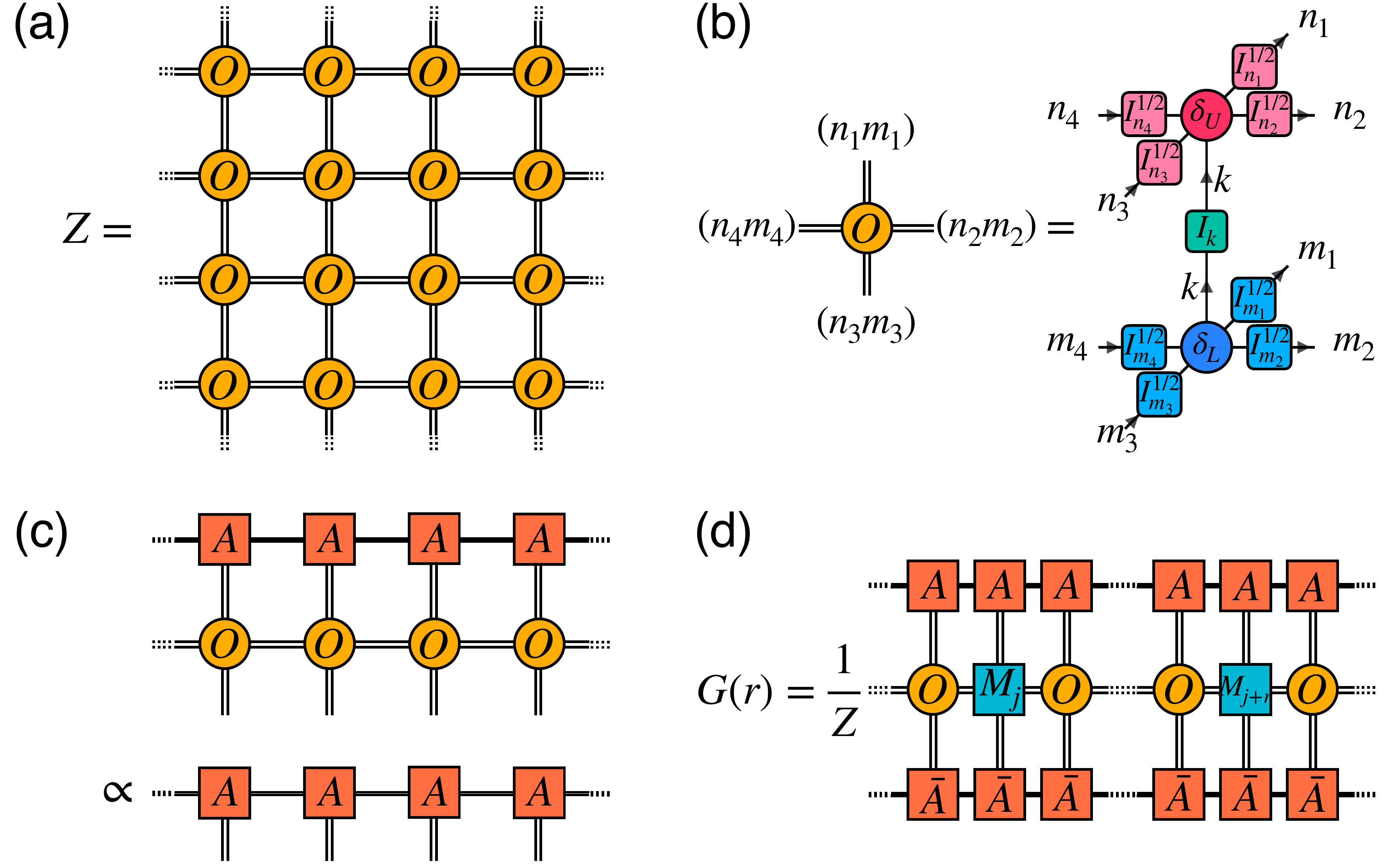}
\caption{ (a) Tensor network representation of the partition function. (b)
The construction of the local tensor $O$ in the partition function. (c)
Eigen-equation for the fixed-point uMPS $|\Psi(A)\rangle$ of the 1D quantum
transfer operator $T$. (d) Two-point correlation function represented by
contracting a sequence of channel operators. }
\label{fig:tensor}
\end{figure}

In the tensor network framework, the partition function is expressed as a
contraction of local tensors defined on the original square lattice, given
by
\begin{eqnarray}
Z &=&\prod_{i}\iint \frac{\mathrm{d}\theta _{i}\mathrm{d}\varphi _{i}}{%
\left( 2\pi \right) ^{2}}\prod_{\langle i,j\rangle }\mathrm{e}^{\beta
J_{1}\cos (\theta _{i}-\theta _{j})}  \notag \\
&&\times \mathrm{e}^{\beta J_{2}\cos (\varphi _{i}-\varphi _{j})}\mathrm{e}%
^{-\beta K\cos (2\theta _{i}-2\varphi _{i})},
\end{eqnarray}%
where $\beta =1/T$ is the inverse temperature. To obtain its tensor network
representation, we apply a duality transformation that maps the phase
variables on each lattice site to the number indices on the nearest-neighbor
links. Such a map is achieved by the character decomposition $\mathrm{e}%
^{x\cos \theta }=\sum_{n=-\infty}^{\infty }I_{n}(x)\mathrm{e}^{in\theta }$
for each Boltzmann factor, where $I_{n}(x)$ is the modified Bessel function
of the first kind. Then the partition function is represented as
\begin{eqnarray}
Z &=&\prod_{i}\iint \frac{\mathrm{d}\theta _{i}\mathrm{d}\varphi _{i}}{%
\left( 2\pi \right) ^{2}}\prod_{l\in \mathcal{L}%
}\sum_{n_{l},m_{l},k_{l}}I_{n_{l}}(\beta J_{1})I_{m_{l}}(\beta J_{2})  \notag
\\
&&\times I_{k_{l}}(-\beta K)\mathrm{e}^{in_{l}(\theta _{i}-\theta _{j})}%
\mathrm{e}^{im_{l}(\varphi _{i}-\varphi _{j})}\mathrm{e}^{i2k_{l}(\theta
_{i}-\varphi _{i})},
\end{eqnarray}%
where $n_{l}$ ($m_{l}$) runs over every link on the upper (lower) layer, and
$k_{l}$ corresponds to every vertical link between $\theta _{i}$ and $%
\varphi _{i}$. By integrating out all the phase degrees of freedom, the
partition function is transformed into a double tensor network as shown in
Fig.~\ref{fig:tensor}(a)
\begin{equation}
Z=\mathrm{tTr}\prod_{i}O_{n_{1}m_{1},n_{2}m_{2}}^{n_{3}m_{3},n_{4}m_{4}}(i),
\label{eq:TN}
\end{equation}%
where \textquotedblleft tTr" denotes the tensor contraction over all
auxiliary links. The details are given in Supplemental Materials. As
displayed in Fig.~\ref{fig:tensor}(b), each local tensor $O$ is defined by
\begin{eqnarray}
O_{n_{1}m_{1},n_{2}m_{2}}^{n_{3}m_{3},n_{4}m_{4}} &=&\sum_{k}\left(
\prod_{l=1}^{4}I_{n_{l}}(\beta J_{1})I_{m_{l}}(\beta J_{2})\right) ^{1/2}
\notag \\
&&\times I_{k}(\beta K)\delta _{n_{1}+n_{2}+2k}^{n_{3}+n_{4}}\delta
_{m_{1}+m_{2}}^{m_{3}+m_{4}+2k},
\end{eqnarray}%
where the inter-layer $k$ indices are summed over and the corresponding
intra-layer $m_{l}$ and $n_{l}$ indices are grouped together. The global $%
U(1)$ invariance of the bilayer model is encoded in each local tensor: $%
O_{n_{1}m_{1},n_{2}m_{2}}^{n_{3}m_{3},n_{4}m_{4}}\neq 0$ only if $%
n_{1}+m_{1}+n_{2}+m_{2}=n_{3}+m_{3}+n_{4}+m_{4}$. Since the expansion
coefficients in the Bessel function $I_{n}(x)$ decrease exponentially as
increasing $n$, an accurate truncation can be performed on the virtual
indices of the local tensors.

In the tensor-network approach, the row-to-row transfer matrix composed of
an infinite row of $O$ tensors is a 1D quantum transfer operator, whose
logarithmic form gives rise to a 1D quantum model with complex spin-spin
interactions. Under such a correspondence, the finite-temperature properties
of the 2D statistical problem are exactly mapped into a 1D quantum model at
zero temperature. In the thermodynamic limit, the value of the partition
function is determined by the dominant eigenvalues of the transfer operator,
whose eigen-equation sketched in Fig.~\ref{fig:tensor}(c) is
\begin{equation}
T|\Psi (A)\rangle =\Lambda _{\max }|\Psi (A)\rangle ,  \label{eq:TM}
\end{equation}%
where $|\Psi (A)\rangle $ is the leading eigenvector represented by uniform
matrix product states (uMPS) consisting of local $A$ tensors\cite%
{Zauner_Stauber_2018}. This eigen-equation can be accurately solved by the
algorithm of variational uniform matrix product states\cite%
{VUMPS,Fishman_2018,Laurens_Haegeman_2019,Laurens_Bram_2019}, and the
largest eigenvector $|\Psi(A)\rangle $ corresponds to the fixed-point
solution. The precision of this approximation is controlled by the auxiliary
bond dimension $D$ of the local $A$ tensors.

From the fixed-point uMPS for the 1D quantum transfer operator, various
physical quantities can be estimated accurately. As far as the phase
transitions are concerned, the quantum entanglement entropy is the most
efficient measure\cite{Vidal_2003,Pollmann_2009}, which can be directly determined via the
Schmidt decomposition of $|\Psi (A)\rangle $: $S_{E}=-\sum_{\alpha
=1}^{D}s_{\alpha }^{2}\ln s_{\alpha }^{2}$, where $s_{\alpha }$ are the
singular values. And the two-point correlation function of the local
observable $h_{i}$ defined by $G(r)=\langle h_{j}h_{j+r}\rangle $ can be
evaluated by the trace of an infinite sequence of channel operators
containing two local impurity tensors $M_j$ and $M_{j+r}$, as shown in Fig.~%
\ref{fig:tensor}(d). The details can be found in Supplementary Materials.

\begin{figure}[tbp]
\centering
\includegraphics[width=0.45\textwidth]{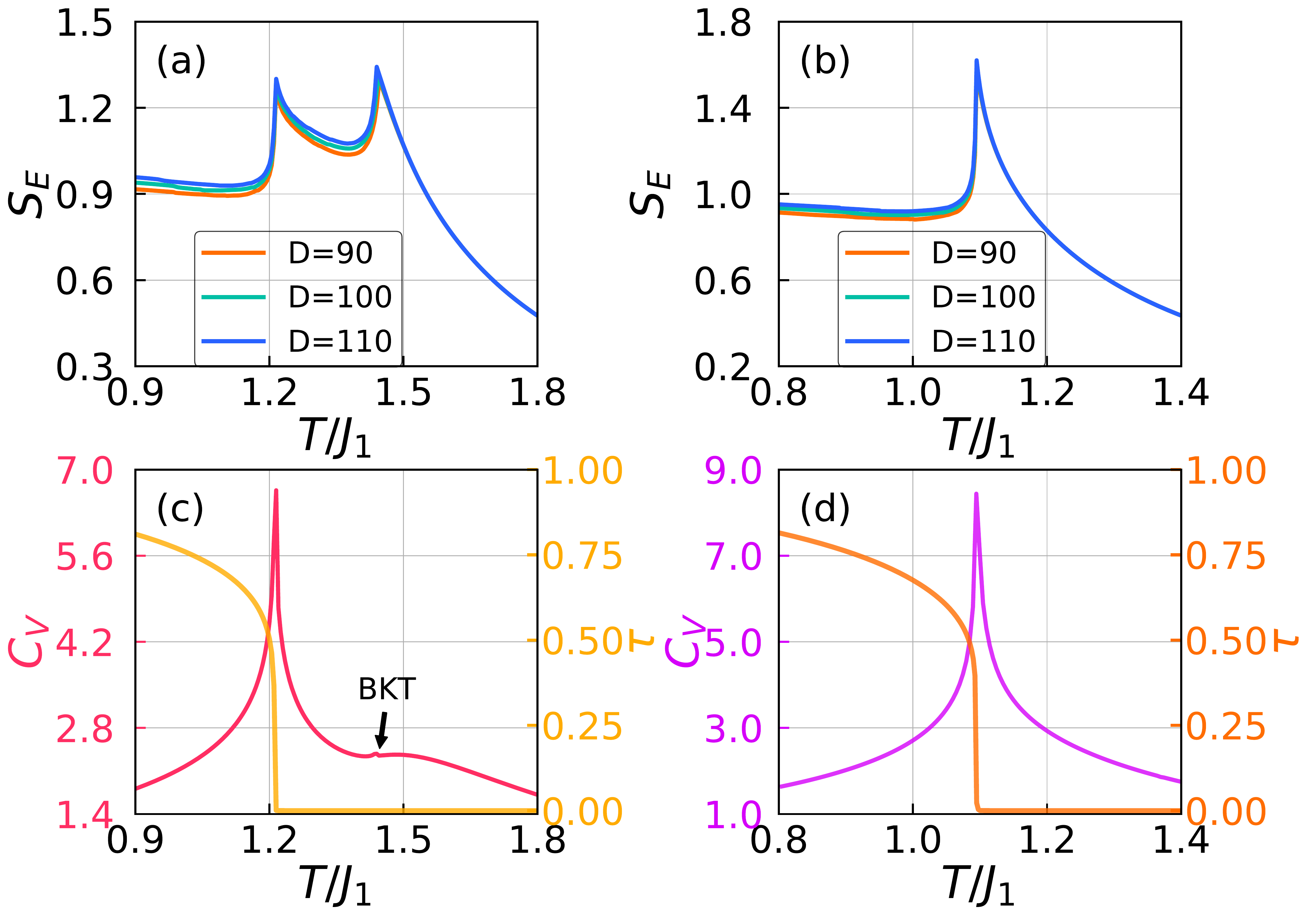}
\caption{ (a) and (b) The entanglement entropy as a function of temperature
for $J_2/J_1=1.5$ and $J_1=J_2$ with $K=0.5J_1$. (c) and (d) The specific
heat and the local Ising order parameter along $J_2/J_1=1.5$ and $J_1=J_2$,
respectively. }
\label{fig:ee-cv}
\end{figure}

\textit{Phase Diagram}. -Since the inter-layer coupling is always relevant,
the structure of the complete phase diagram is \textit{independent} of its
value, so we simply choose a practical value $K/J_{1}=0.5$. Importantly we
have noticed that the entanglement entropy $S_{E} $ of the fixed-point uMPS
for the 1D quantum transfer operator exhibits singularity, which provides an
accurate criterion to determine the transition points. To obtain the phase
diagram, we have to numerically calculate the entanglement entropy under a
wide range of intra-layer coupling ratios $J_{2}/J_{1}$. In Fig.~\ref%
{fig:ee-cv}(a), the entanglement entropy along $J_{2}/J_{1}=1.5$ develops
two sharp peaks at $T_{c1}\simeq 1.21J_{1}$ and $T_{c2}\simeq 1.44J_{1}$,
respectively. When $J_{2}$ approaches $J_{1}$, these two peaks merge
together, leading to a single peak at $T_{\ast}\simeq 1.095J_{1}$ as shown
in Fig.~\ref{fig:ee-cv} (b). These peak positions are nearly unchanged under
the bond dimensions $D=90,100,110$. So the phase boundaries can be
determined with high precision and the complete phase diagram is displayed
in Fig.~\ref{fig:phase}(a).

In order to gain insight into the essential physics of different phases, we
calculate the specific heat. Within the tensor-network framework, the
internal energy per site is calculated as
\begin{equation*}
u=-2J_{1}\langle \mathrm{e}^{i(\theta _{j}-\theta _{j+1})}\rangle
-2J_{2}\langle \mathrm{e}^{i(\varphi _{j}-\varphi _{j+1})}\rangle +K\langle
\mathrm{e}^{i(2\theta _{j}-2\varphi _{j})}\rangle ,
\end{equation*}%
and the specific heat is obtained by $C_{V}=\partial u/\partial T$. As shown
in Fig.~\ref{fig:ee-cv}(c), along the line $J_{2}/J_{1}=1.5$, the specific
heat exhibits a logarithmic divergence at $T_{c1}$ but a small bump around $%
T_{c2}$. However, for $J_{2}/J_{1}=1$, a single logarithmic singularity is
observed at $T_{\ast }$ as displayed in Fig.~\ref{fig:ee-cv}(d). The
logarithmic specific heat at the lower temperature reminds us of a 2D Ising
phase transition with a $\mathbb{Z}_{2}$ symmetry breaking, while the small
bump at the higher temperature indicates the nearby BKT transition.

At low temperatures, since the relative phase field $\sigma_{i}\equiv%
\theta_{i}-\varphi_{i}$ is reduced to a $\mathbb{Z}_2$ variable, a local
inter-layer Ising order parameter can be defined by $\tau =\langle \sin
\sigma _{i}\rangle$. As shown in Fig.~\ref{fig:ee-cv}(c) and (d), $\tau $ is
finite below $T_{c1}$, indicating that the phase lock occurs between the
upper and lower layers. When $J_{2}/J_{1}=1$, the Ising transition coincides
with the BKT transition exactly at the multi-critical point $P$, where there
is an interplay between the Ising and BKT degrees of freedom at the
microscopic level and exhibits a new universality class of critical
properties with emerged supersymmetry\cite{Huijse_2015}.

\textit{Correlation functions and spin stiffness}. -To further explore the
nature of the intermediate temperature phase, we calculate the two-point
correlation functions of the XY spins and nematic spins, which represent the
integer vortices and half-integer vortices variables in the bilayer system,
respectively. The results are summarized in Table.1.

\begin{table}[tbp]
\caption{Properties of correlation functions in the different phases of the
phase diagram in Fig.\protect\ref{fig:phase}.}\centering
\begin{tabular}[t]{cllll}
\hline
\toprule & disordered & quasi-LRO-1 & quasi-LRO-2 & ordered \\ \hline
\midrule $\!\langle \mathrm{e}^{i(\varphi_i-\varphi_j)}\rangle\,$ & $\sim%
\mathrm{e}^{-r/\xi_\varphi}$ & $\sim\mathrm{e}^{-r/\xi_\varphi}$ & $\sim
r^{-\eta_\varphi}$ & $\sim r^{-\eta_\varphi}$ \\
$\!\langle \mathrm{e}^{i2(\varphi_i-\varphi_j)}\rangle\,$ & $\sim\mathrm{e}
^{-r/\xi_{2\varphi}}$ & $\sim r^{-\eta_{2\varphi}}$ & $\sim
r^{-\eta_{2\varphi}}$ & $\sim r^{-\eta_{2\varphi}}$ \\
$\!\langle \mathrm{e}^{i(\theta_i-\theta_j)}\rangle\,$ & $\sim\mathrm{e}
^{-r/\xi_\theta}$ & $\sim r^{-\eta_\theta} $ & $\sim\mathrm{e}%
^{-r/\xi_\theta}$ & $\sim r^{-\eta_\theta} $ \\
$\!\langle \mathrm{e}^{i2(\theta_i-\theta_j)}\rangle\,$ & $\sim\mathrm{e}
^{-r/\xi_{2\theta}}$ & $\sim r^{-\eta_{2\theta}}$ & $\sim
r^{-\eta_{2\theta}} $ & $\sim r^{-\eta_{2\theta}} $ \\
$\!\langle \mathrm{e}^{i(\theta_i-\varphi_j)}\rangle\,$ & $\sim\mathrm{e}
^{-r/\xi_{\theta\varphi}}$ & $\sim\mathrm{e} ^{-r/\xi_{\theta\varphi}}$ & $%
\sim\mathrm{e}^{-r/\xi_{\theta\varphi}}$ & $\sim r^{-\eta_{\theta\varphi}}$
\\
$\!\langle \mathrm{e}^{i(\sigma_i-\sigma_j)}\rangle\,$ & $\sim\mathrm{e}
^{-r/\xi_{\sigma}}$ & $\sim\mathrm{e} ^{-r/\xi_{\sigma}}$ & $\sim\mathrm{e}%
^{-r/\xi_\sigma}$ & $\sim$ const. \\ \hline
\bottomrule &  &  &  &
\end{tabular}%
\end{table}

For $J_2/J_1>1$, the spin-spin correlation function of the lower layer $%
G_{\varphi }(r)$ starts to decay algebraically at $T_{c2}$ as the
temperature decreases. When approaching $T_{c2}$ from above, the spin
correlation length $\xi _{\varphi }$ is well-fitted by an exponentially
divergent form
\begin{equation}
\xi (T)\propto \exp (\frac{b}{\sqrt{T-T_{C}}}),\quad T\rightarrow T_{C}^{+},
\label{eq:BKT_cl}
\end{equation}
where $b$ is a non-universal positive constant. This is the characteristic
feature of the BKT transition. Below $T_{c1}$, the spin-spin correlation
functions of both the intra-layer $G_{\theta }(r)$ and the inter-layer $%
G_{\theta\varphi}(r)$ exhibit the algebraic behavior, implying the
vortex-antivortex bindings in both intra-layers and inter-layers, a fully
phase-coherent state of the Cooper pairs in the bilayer system.

\begin{figure}[tbh]
\centering
\includegraphics[width=0.45\textwidth]{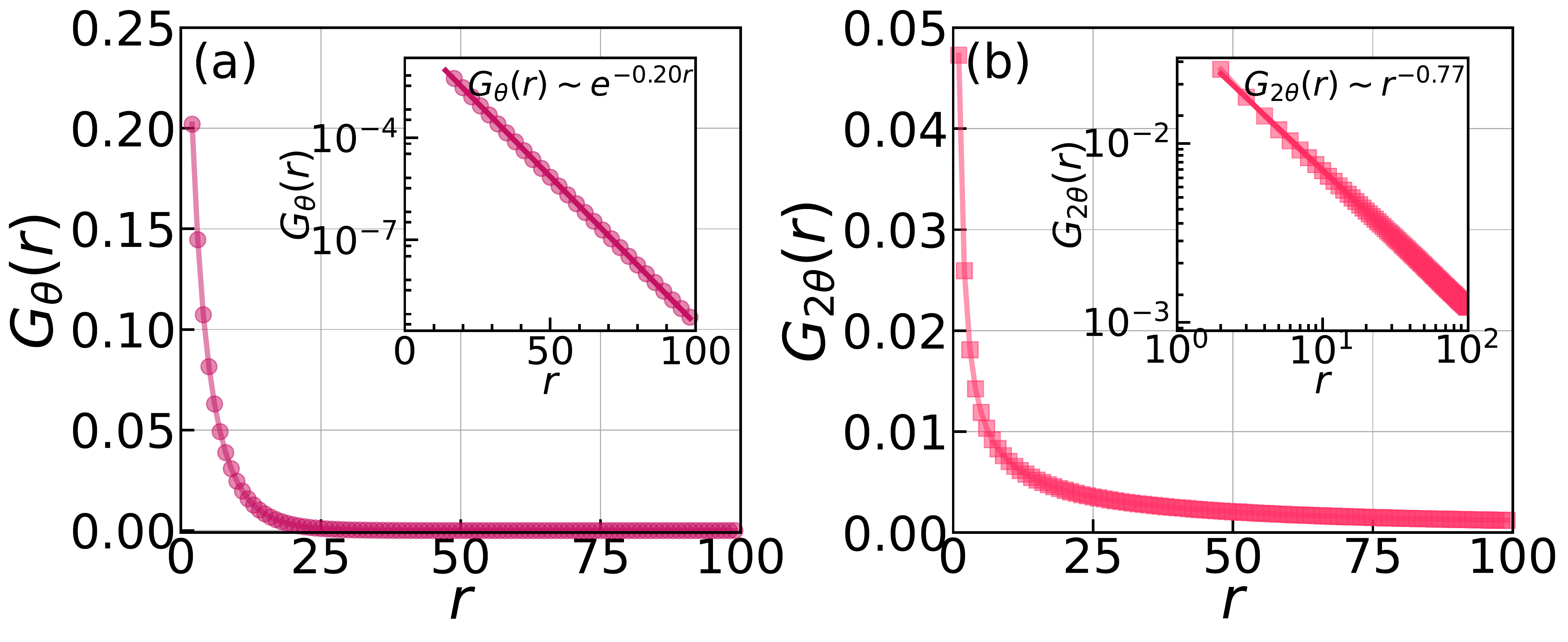}
\caption{ The properties of the quasi-LRO phase-2 when $J_2/J_1=1.5$, $%
K/J_1=0.5$, and $T/J_1=1.3$. (a) The correlation function of the XY spins
shows an exponential decay. (b) The correlation function of the nematic
spins exhibits a power law decay.}
\label{fig:correlation}
\end{figure}

When we focus on the quasi-LRO-2 phase, the spin-spin correlation function $%
G_{\varphi }(r)$ in the lower layer decays algebraically, while in the upper
layer it is the correlation function of the nematic spins $G_{2\theta }(r)$
that exhibits an algebraic behavior, instead of the correlation function of
the XY spins $G_{\theta }(r)$
\begin{eqnarray}
&&G_{\theta }(r)=\langle e^{i\left( \theta _{j}-\theta _{j+r}\right)}\rangle
\sim \mathrm{e} ^{-r/\xi_\theta},  \notag \\
&&G_{2\theta }(r)=\langle e^{i\left( 2\theta_{j}-2\theta_{j+r}\right)
}\rangle \sim r^{-\eta_{2\theta}}.
\end{eqnarray}
For a given value of $J_2/J_1=1.5$ and $T/J_1=1.3$, the comparison between
the spin-spin correlation function and nematic correlation function is
displayed in Fig.~\ref{fig:correlation}(a) and (b). Such a behavior
indicates that the integer vortices in the upper layer are fractionalized
into half-integer vortex pairs due to the presence of the inter-layer
squared cosine interaction. Since the half-integer vortices are point-like
topological defects about which the phase angles of spins wind by $\pi$,
each pair of half-vortices should be connected by a topological string
across which spins are antiparallel. Because the integer vortex-antivortex
pairs with quasi-LRO are regarded as the phase condensation of the Cooper
pairs in 2D, the half-integer vortex pairs with quasi-LRO can be regarded as
the condensation of pairs of the Cooper pairs in the absence of the phase
coherence among the Cooper pairs\cite{Doucot_2002,Babaev_2004}. Such a
phenomenon is just the characteristics of the charge-4e superconductivity%
\cite{Berg_2009,Yao_2017,Fu_2021}.

\begin{figure}[tbh]
\centering
\includegraphics[width=0.45\textwidth]{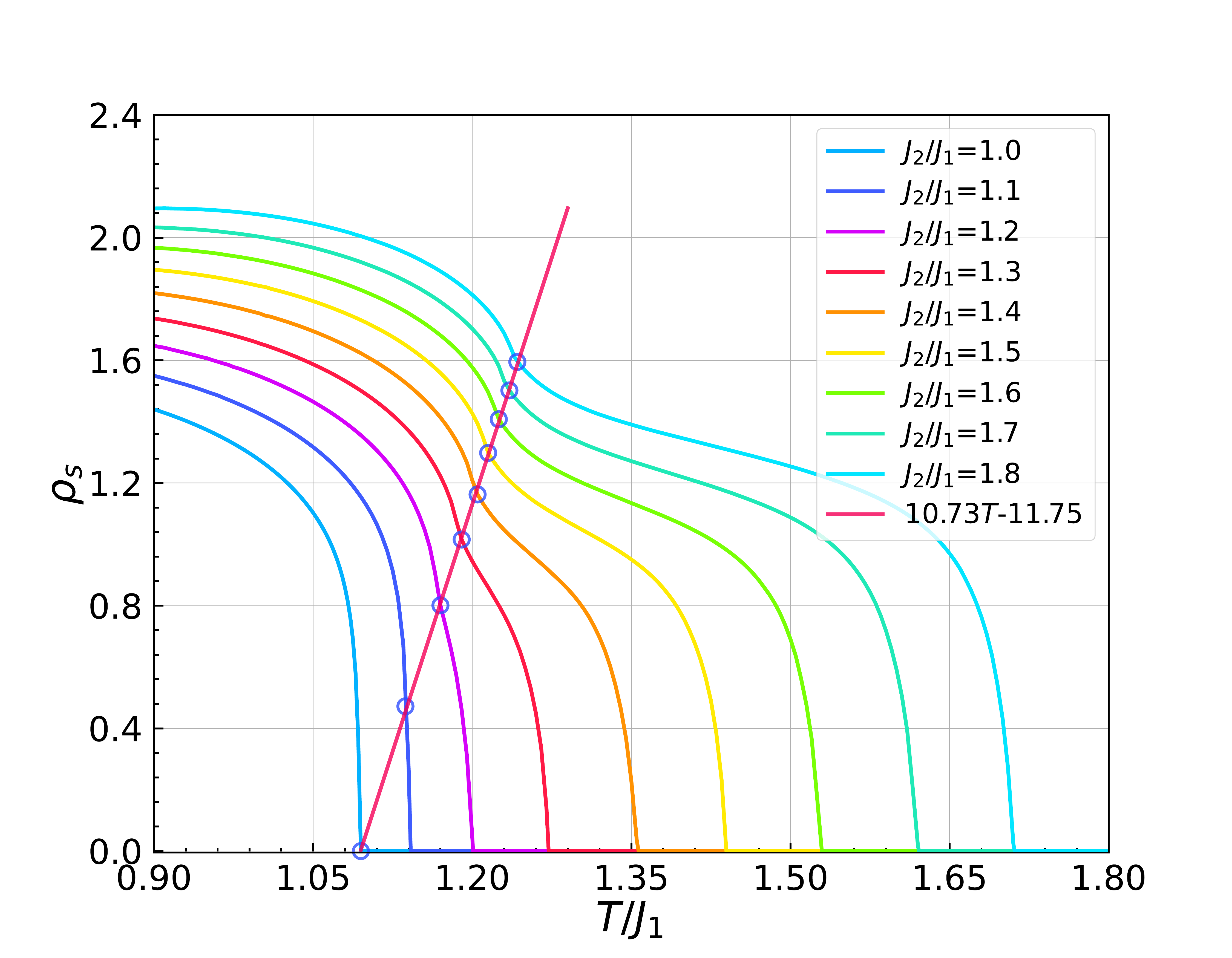}
\caption{The spin stiffness as a function of temperature for given values of
$J_2/J_1$. The inter-layer coupling is chosen as $K=0.5J_1$ and the bond
dimension of the local tensor is $D=110$. The red straight line indicates
the temperature of the Ising transition.}
\label{fig:stiffness}
\end{figure}

To access the superfluid response of the bilayer system, we calculate the
spin stiffness or the helicity modulus defined by the second derivative of
the free energy density with respect to a twist $v$ along a given direction\cite{Fisher_1973,Nelson_1977},
$\rho_s=\frac{\partial^2 f}{\partial ^2 v}|_{v=0}$. The twist needs to be
taken in a way that respects the joint $U(1)$ symmetry of the coupled
bilayer, and the spin stiffness is expressed in terms of two-point functions
within the framework of tensor network methods\cite{Vanderstraeten_2015,Vanderstraeten_2016}. Since the process is more technical, the details are given in the Supplementary Materials. The jump of
spin stiffness should be altered from the BKT predictions $%
\rho_s/T_{BKT}=2/\pi$ due to the emergence of half vortices\cite%
{Hubscher_2013}. In Fig.~\ref{fig:stiffness}, the numerical spin stiffness
as a function of temperature is shown for $J_2/J_1=1.0\sim 1.8$ with the
inter-layer coupling $K/J_1=0.5$. It can be seen that the spin stiffness
starts to dramatically increase from zero around the BKT transition
temperature $T_{c2}$. When the temperature decreases, a cusp point forms in
the further increase of the spin stiffness, corresponding to the Ising phase
transition $T_{c1}$ precisely. Surprisingly, the cusp points for given
values of $J_2/J_1$ sit on a straight line, which is a key experimental
feature of the presence of the Ising phase transition within the
superconducting phase.

\textit{Conclusion}. -We have used the tensor-network methods to study the
bilayer system of two-coupled 2D XY spin models. The global finite
temperature phase diagram has been accurately determined. It has been found
that, as the temperature decreases, the BKT transition always happens above
the phase locking of the bilayer system, which corresponds to an inter-layer
Ising long-range order. More importantly, for two inequivalent coupled
bilayers, there exists an intervening unlocked phase, where the half-integer
vortex pairs form in one layer with the smaller intra-layer coupling,
coexisting with the integer vortex-antivortex pairs in the other layer. When
a weak direct Josephson coupling is also present, we have further proved
that the Ising phase transition below the BKT transition survives and the
main results of this work are still valid, because two local minima always
exist to lock the phase fields of the upper and lower layers.

Recently a new family of superconductors ACa$_2$Fe$_4$As$_4$F$_2$
(A=K,Rb,Cs) has synthesized\cite{Wang_2016}, and these compounds can be
viewed as an intergrowth of AFe$_2$As$_2$ and CaFeAsF layers. The transport
and magnetic measurements on single crystals of CsCa$_2$Fe$_4$As$_4$F$_2$
showed a large resistivity anisotropy that tends to increase with decreasing
temperature, and the 2D superconducting fluctuations have been observed\cite%
{Cao-2019}. The evolution of the in-plane penetration depth shows an
inflection point around $10$ K, indicating that a potentially "magnetic"
phase appears but does not compete with superconductivity\cite{Blundell-2018}%
. These features may be related to the formation of the inter-layer Ising
long-range order and the manifestation of the phase coherence of pairs of
Cooper pairs revealing a cusp point in the spin stiffness. Therefore, these
compounds are good candidate systems to explore the charge-4e
superconductivity.

\textbf{Acknowledgments.} The authors are indebted to Qi Zhang for his
stimulating discussions. The research is supported by the National Key
Research and Development Program of MOST of China (2017YFA0302902).

\bibliography{reference}

\end{document}